\documentclass[twocolumn,prb,amsmath,amssymb,superscriptaddress]{revtex4-1}
\usepackage[utf8]{inputenc}
\usepackage{amsfonts,amsmath,amssymb}
\usepackage{graphicx}
\usepackage{bm} 
\usepackage{mathrsfs}
\usepackage{subfigure}
\usepackage{textcomp}
\usepackage{hyperref}
\usepackage[flushleft]{threeparttable}
\usepackage[per-mode=symbol]{siunitx}
\usepackage[version=3]{mhchem}

\def\qr{{\bf r}}                                   

\begin{document}

\title{Suppression of polaron self-localization by correlations}

\author{Lilith Zschetzsche}
\author{Robert E. Zillich}
\affiliation{Institute for Theoretical Physics, Johannes Kepler University Linz, Altenberger Straße 69, 4040 Linz, Austria}

\date{\today}

\begin{abstract}
We investigate self-localization of a polaron in a homogeneous Bose-Einstein condensate in one dimension.
This effect, where an impurity is trapped by the deformation that it causes in the surrounding Bose gas,
has been first predicted by mean field calculations, but has not been seen in experiments.
We study the system in one dimension, where, according to the mean field approximation,
the self-localization effect is particularly robust, and present for arbitrarily weak impurity-boson interactions.
We address the question whether self-localization is a real effect
by developing a variational method which incorporates impurity-boson correlations non-perturbatively 
and solving the resulting inhomogeneous correlated polaron equations. We find that correlations
inhibit self-localization except for very strongly repulsive or attractive impurity-boson
interactions. Our prediction for the critical interaction strength for self-localization agrees with a sharp drop
of the inverse effective mass to almost zero found in quantum Monte Carlo simulations of polarons in one dimension.

\end{abstract}

\maketitle

\section{Introduction} 

The original Bose polaron problem concerns an electron in a solid which is dressed by
small distortions of the crystal lattice and was modelled by Fr\"ohlich \cite{froehlich54}.
Another type of polaron is formed by an electron or impurity atom in superfluid $^4$He. This
problem has long been studied\cite{tabbertJLTP97} and later extended to molecular impurities and
impurity aggregates in $^4$He, which lead to a new type
of low-temperature spectroscopy of molecules~\cite{toenniesAngChem04,slenczkaToennies}.
More recently, polarons of mobile impurities have been experimentally realized
in ultracold Bose gases\cite{catani_quantum_2012,huPRL16,jorgensenPRL16}.

For electrons in ionic solids~\cite{landau33} and also in superfluid $^4$He~\cite{jortnerJCP65}
a mechanism for self-localization, or self-trapping, was proposed\cite{emin}. Self-localization
implies that, even in the absence of an external trap potential,
the impurity probability density $\rho(\qr_0)$ is not uniform but trapped by the distortion of the
density of phonons or He atoms created by the impurity itself.
In Refs.~ \onlinecite{cucchiettiPRL06,kalasPRA06} based on the mean field (MF) approach self-localization has also been predicted
for polarons in a Bose-Einstein condensate. According to
Cuccietti et al.\cite{cucchiettiPRL06} 
a polaron in a three-dimensional homogeneous Bose gas self-localizes above a critical
impurity-boson interaction strength, while below it the polaron ground state is homogeneous. This would imply a
phase transition to a translation symmetry breaking ground state.  Subsequently,
other works have also predicted this effect, e.g.\ for neutral polarons, again using the MF approximation
\cite{sachaTimmermansPRA06,brudererEPL08,santamoreNJP11,blinovaPRA13,liSciRep13}, including finite temperature
\cite{boudjemaa_self-localized_2014}, and also with other methods such as path integrals
\cite{temperePRB09,novikov_variational_2010}. Also ionic polarons
\cite{casteels_polaronic_2011} and angular polarons
\cite{liPRA17angularselfloc} have been predicted to self-localize.
However, other works have not seen evidence of  self-localization in three dimensions \cite{li_variational_2014,ardilaPRA15,hahnPRB18}, nor
has it been observed experimentally. This raises the question whether self-localization is a methodological
artifact or a real effect.

In one dimension the MF approximation predicts a self-trapped polaron regardless of the strength of the impurity-boson
interaction\cite{brudererEPL08}. 
Exact quantum Monte Carlo simulations \cite{parisi_quantum_2017} indeed predict an
essentially divergent polaron effective mass above a certain impurity-boson interaction strength,
i.e. the polaron becomes immobile, which would be consistent with self-localization for strong interactions. Conversely,
Ref.~\onlinecite{grusdtNJP17} found a finite effective mass for attractive impurity-boson interaction, using the same Monte Carlo method for similar boson-boson interaction strengths but smaller mass ratio. Indirect measurements of Bose polarons in one dimension gave
an even lower effective mass~\cite{catani_quantum_2012}.

The goal of this work is to check if the self-localized ground state predicted by the MF
approximation is a real effect or an artifact of the uncorrelated Hartree ansatz of MF.
To check this, we take a crucial step beyond the Hartree ansatz by incorporating
impurity-boson correlations in a non-perturbative way, while treating the weakly interacting Bose background still
in the MF approximation, thus omitting boson-boson correlations. We note that the perturbative treatment of correlations (then usually referred to as quantum fluctuations) has been shown to lead to corrections to the density $\rho(x_0)$ of a self-localized impurity in one
dimension \cite{sachaTimmermansPRA06} but still preserves self-localization. In this work we show that
with a non-perturbative treatment of impurity-boson correlations impurity self-localization happens only
for very strongly attractive or repulsive impurity-boson interactions.


\section{Theory and Method} 

The Hamiltonian of one impurity and $N$ bosons in one dimension is given by
\begin{align}
 H &= -{\hbar^2\over 2M}{\partial^2\over \partial x_0^2} - {\hbar^2\over 2m}\sum_{i=1}^N{\partial^2\over \partial x_i^2} +\sum_{i=1}^N U(x_0-x_i) \nonumber\\
 &+ \lambda_\text{BB}\sum_{i<j} \delta(x_i-x_j)
 \label{eq:H}
\end{align}
consisting of the kinetic energy of the impurity, the kinetic energy of the bosons, the impurity-boson interaction
and the boson-boson interaction.
The boson-boson interaction is modelled by a contact potential with strength $\lambda_\text{BB}$, which is
related to the scattering length $a_\text{BB}$ by $\lambda_\text{BB} = \frac{-2\hbar^2}{a_\text{BB}m}$ \cite{Bloch,pitaevskiibook}.
The impurity-boson interaction is modelled by a finite range potential, for which we choose
a Gaussian,
$U(x)={U_0\over 2\sigma_U^2}\exp\left[-{x^2\over\sigma_U^2}\right]$, 
characterized by the strength and width parameters $U_0$
and $\sigma_U$.

The MF approach is usually derived in a variational formulation, with
the Hartree ansatz wave function for one impurity in a bath of $N$ bosons:
\begin{align}
    \Psi_{\rm MF} = \eta(x_0)\prod\limits_{i=1}^{N}\psi(x_i).
\label{eq:PsiMF}
\end{align}
This wave function does not account for the correlations caused by the interactions, e.g.\ the
decrease of the probability $|\Psi(x_0,\dots,x_i,\dots)|^2$ if a boson at $x_i$ is close to
a repulsive impurity at $x_0$. The optimization of $\Psi_{\rm MF}$ leads to one-body equations with
effective potentials, the ``mean fields''.
The uncorrelated MF ansatz \eqref{eq:PsiMF} can be expected to be
a poor approximation of the true many-body wave function if impurity-boson interactions are strong (but our results show it is a poor approximation for weak interaction as well). Therefore,
we generalize the ansatz by replacing the boson one-body functions
$\psi(x_i)$ with impurity-boson {\it pair correlation} functions $f(x_0,x_i)$:
\begin{align}
    \Psi = {1\over\Omega^{N/2}}\eta(x_0)\prod\limits_{i=1}^{N}f(x_0,x_i), 
\label{eq:Psi}
\end{align}
where it turns out to be convenient to introduce a prefactor including the normalization volume $\Omega$.
This is a Jastrow-Feenberg ansatz wave function \cite{Feenberg} but limited to impurity-boson correlations. We refer to it as the inhomogeneous correlated polaron
(inh-CP) ansatz.

If the ground state is assumed homogeneous, i.e.\ translationally invariant like the Hamiltonian, the ansatz
\eqref{eq:Psi} simplifies to
\begin{align}
    \Psi_\text{hom} = {1\over\Omega^{(N+1)/2}}\prod\limits_{i=1}^{N}f_\text{hom}(x_0-x_i),
\label{eq:Psihom}
\end{align}
which was studied already by Gross~\cite{grossAnnPhys62c}.
Of course, we cannot make this assumption of translational invariance
if we want to study the possible symmetry breaking by self-localization
of the impurity. But the homogeneous correlated polaron (hom-CP) ansatz \eqref{eq:Psihom} will still be useful: if
self-localization is indeed energetically favorable, the energy difference between
the inh-CP and the hom-CP result is the energy gained by forming a self-localized
ground state.


Our ansatz (\ref{eq:Psi}) includes impurity-boson correlations but still treats the (weakly interacting) Bose background in the MF approximation, as it does not include boson-boson correlations. Since we take only one step beyond the MF approach, this allows for a comprehensible comparison between our method and the MF approach.
Impurities immersed in a strongly interacting Bose liquid like $^4$He, however, require the inclusion of boson-boson correlations.
Optimizing such a full Jastrow-Feenberg ansatz leads to the hypernetted-chain Euler-Lagrange
method\cite{KroTrieste,QMBT00Polls}. The method and its time-dependent generalization have been used extensively
to study impurities in $^4$He \cite{paper4theoretikum,effmass98,zillichJCP10b}.

Before deriving equations for $\eta(x_0)$ and $f(x_0,x_i)$ from the Ritz' variational principle, we
need an expression for the energy functional $E = \langle \Psi \vert H \vert \Psi \rangle$, where we assume normalization of the wave function, $\langle \Psi \vert \Psi \rangle=1$.
The 4 terms in the Hamiltonian (\ref{eq:H}) lead to the following 4 terms in $E$:
\begin{align}
  E&=\frac{\hbar^2}{2M} \int\! \mathrm{d}\mathbf{x}\, \left( {\partial \Psi\over\partial x_0} \right)^2 \nonumber\\
   &+ N \frac{\hbar^2}{2m}\int\! \mathrm{d}\mathbf{x}\, \left( {\partial \Psi\over\partial x_1} \right)^2
  + N \int\! \mathrm{d}\mathbf{x} \ \Psi^2 \ U(x_0- x_1) \nonumber\\
   &+ \frac{N(N-1)}{2} \int\! \mathrm{d}\mathbf{x} \ \Psi^2 \ \lambda_\text{BB} \ \delta(x_1- x_2), \label{Equ:E_corr}
\end{align}
where $\mathrm{d}\mathbf{x} = dx_0 \, dx_1 \, ... \, dx_N$, and $\Psi$ is the correlated polaron ansatz (\ref{eq:Psi}).
Owing to the star-shaped correlation structure, where the impurity is correlated with all bosons but the bosons are not
correlated between themselves, most of the $N+1$ integrals in $E$ factorize and yield $\int d x_1' \, f( x_0, x_1')^2$.
We abbreviate this partially integrated correlation function
\begin{align}
    \bar{f}( x_0)\equiv \Omega^{-1}\! \int d x_1' \ f( x_0, x_1')^2.
    \label{eq:barf}
\end{align}
We obtain the energy functional
\begin{widetext}
\begin{align}
E \ =& \ \frac{\hbar^2}{2M} \ \Biggl\{ \int\!d x_0\, \left( \frac{\partial \eta( x_0)}{\partial  x_0} \right) ^2  \bar{f}( x_0)^N
 - \frac{N}{\Omega} \int\! dx_0dx_1\, \eta( x_0)^2\, \bar{f}(x_0)^{N-1} f(x_0, x_1)\,
  \frac{\partial^2 f(x_0, x_1)}{\partial  x_0^2} \nonumber\\
& \hspace{1.05cm} - \frac{N(N-1)}{\Omega^2} \int\! dx_0\, \eta( x_0)^2 \, \bar{f}( x_0)^{N-2} \left[ \int\! d x_1 \, f(x_0, x_1)\,
  \frac{\partial f(x_0, x_1)}{\partial  x_0} \right]^2 \Biggr\} \nonumber\\
&+ \frac{\hbar^2}{2m} \frac{N}{\Omega} \int\! dx_0 dx_1\, \eta( x_0)^2 \, \bar{f}( x_0)^{N-1} 
  \left( \frac{\partial f(x_0, x_1)}{\partial x_1} \right) ^2
 + \frac{N}{\Omega} \int\! dx_0dx_1 \, \eta(x_0)^2 \, \bar{f}(x_0)^{N-1}\, f(x_0, x_1)^2 \, U( x_0, x_1) \nonumber\\
&+ \frac{\lambda_\text{BB}}{2} \ \frac{N(N-1)}{\Omega^2} \int\! dx_0dx_1 \, \eta( x_0)^2 \, \bar{f}(x_0)^{N-2}\, f( x_0, x_1)^4.
\label{Equ:Efull2}
\end{align}
\end{widetext}

In a study of self-localization, we are primarily interested in the impurity density $\rho_\text{I}(x_0)$.
Without an external trapping potential, the impurity density is constant in the absence of self-localization, $\rho_\text{I}(x_0)={1\over\Omega}$,
while in the presence of self-localization $\rho_\text{I}(x_0)$ peaks at a random location $\bar x_0$ \footnote{For numerical reasons,
the impurity self-localizes at $\bar x_0=0$ if at all.} and falls to zero away from $\bar x_0$. Similarly,
the density of the Bose gas $\rho_\text{B}(x_1)$ is constant in the first case, $\rho_\text{B}(x_1)={N\over\Omega}$ ,
while it has a valley/peak for repulsive/attractive impurity-boson interaction in the latter case.
For the correlated polaron ansatz~\eqref{eq:Psi}, the impurity density is given by
\begin{align}
    \rho_\text{I}(x_0) = \int\!dx_1\dots dx_N |\Psi|^2 = \eta(x_0)^2 \bar{f}(x_0)^{N},
    \label{eq:defrhoI}
\end{align}
and the boson density is given by
\begin{align}
    \rho_\text{B}(x_1) &= N\int\!dx_0 dx_2\dots dx_N |\Psi|^2 \nonumber\\
    &= {N\over\Omega} \int\!dx_0\ \eta(x_0)^2 \bar{f}(x_0)^{N-1} f(x_0,x_1)^2,
    \label{eq:defrhoB}
\end{align}
where normalization of the wave function was assumed.

According to the Ritz' variational principle the optimal $\eta(x_0)$ and $f(x_0,x_1)$ are obtained from
minimizing the energy \eqref{Equ:Efull2}, i.e.\ setting its functional derivatives with respect to $\eta(x_0)$ and $f(x_0,x_1)$
to zero. To ensure normalization of the wave function we introduce a Lagrange multiplier
$\lambda$. Hence, we need to optimize the Lagrangian
\begin{equation}
L=E+\lambda \left\lbrace 1-\int d x_0 \ \eta( x_0)^2 \ \bar{f}( x_0)^N \right\rbrace. \label{Equ:Lagrange_mu}
\end{equation}
The inh-CP equations for the general inhomogeneous case
are the coupled Euler-Lagrange equations, formally written as
\begin{align}
	&\hspace{0.3cm}\frac{\delta L}{\delta \eta(x_0)}=0, \label{Equ:funcderiv_eta}\\
	&\frac{\delta L}{\delta f(x_0,x_1)}=0. \label{Equ:funcderiv_f}
\end{align}
Their explicit form is derived in appendix~\ref{app1}, where we show that in the thermodynamic limit $N\to\infty$ and $\Omega\to\infty$
with $\rho={N \over\Omega}$ fixed, we obtain a 1-body equation for the square root of the impurity
density $g(x_0)=\sqrt{\rho_\text{I}(x_0)}$
and a two-body equation for $\tilde{f}( x_0, x_1) \equiv g( x_0) f( x_0, x_1)$:
\begin{widetext}
\begin{align}
    \mu_\text{I} \ g( x_0) =
  &-\frac{\hbar^2}{2M} \ \frac{\partial^2 g( x_0)}{\partial  x_0^2} + V_g(x_0)\, g(x_0)\label{Equ:var_eta_approx3}\\
    \mu_\text{B} \ \tilde{f}( x_0, x_1) =
  & - \frac{\hbar^2}{2M} \frac{\partial^2 \tilde{f}( x_0, x_1)}{\partial { x_0}^2}
    -\frac{\hbar^2}{2m} \frac{\partial^2 \tilde{f}( x_0, x_1)}{\partial { x_1}^2} + V_f(x_0,x_1)\, \tilde{f}( x_0, x_1)
  \label{Equ:var_f_approx3}
\end{align}
with the impurity and boson chemical potential $\mu_\text{I}$ and $\mu_\text{B}$ and the effective one-body and two-body potentials
\begin{align}
  V_g(x_0) &= \frac{\hbar^2}{2M} \rho \int\! d x_1' \left( \frac{\partial f( x_0, x_1')}{\partial  x_0} \right)^2
    + \frac{\hbar^2}{2m} \rho \int\! d x_1' \left( \frac{\partial f( x_0, x_1')}{\partial  x_1'} \right) ^2
    + \rho \int\! d x_1' \, f( x_0, x_1')^2 \, U( x_0, x_1') \nonumber\\
   &+ \lambda_\text{BB}\frac{\rho^2}{2} \int\! d x_1' \, \Bigl( f( x_0, x_1')^4 - 2 \ f( x_0, x_1')^2 +1 \Bigr)
   \label{Equ:Vg}\\
  V_f(x_0,x_1) &=  \frac{\hbar^2}{2M} \frac{1}{g( x_0)} \frac{\partial^2 g( x_0)}{\partial  x_0^2} + U( x_0, x_1)
    + \lambda_\text{BB} \rho \ \frac{\tilde{f}( x_0, x_1)^2}{g( x_0)^2}.
 \label{Equ:Vf}
\end{align}
\end{widetext}
We have cast the two coupled inh-CP equations into the form of a one- and a two-body nonlinear Schr\"odinger equation,
respectively, with effective potentials \eqref{Equ:Vg} and \eqref{Equ:Vf}
that depend on $g(x_0)$ and $\tilde{f}( x_0, x_1)$ itself. Similarly to other nonlinear
Schr\"odinger equations\cite{chinPRE05b},  Eqns.~\eqref{Equ:var_eta_approx3} and \eqref{Equ:var_f_approx3}
can be solved self-consistently by imaginary time propagation, where we always start
the propagation with self-localized trial states, for example the MF ground state.
Details are given in appendix~\ref{app2}.

\section{Results} 

We present results for the Bose polaron ground state in one dimension for
three levels of approximation:
\begin{enumerate}
  \item[a)] solving the full inh-CP equations
    \eqref{Equ:var_eta_approx3} and \eqref{Equ:var_f_approx3},
    derived in this work and based on the ansatz~\eqref{eq:Psi};
  \item[b)] solving the special case of the hom-CP
    equation, derived in Ref.~\onlinecite{grossAnnPhys62c}, based on the
    ansatz~\eqref{eq:Psihom}, that precludes self-localization;
  \item[c)] solving the MF equations, based on the ansatz
    \eqref{eq:PsiMF}, which according to Ref.~\onlinecite{brudererEPL08} always result in self-localization in one dimension.
\end{enumerate}
In all three types of calculations, we use the same Gaussian interaction model.
Following Bruderer et al.\cite{brudererEPL08},
we measure length in units of the healing length $\xi=\hbar/\sqrt{\lambda_\text{BB}\rho m}$ and energy in units
of $E_0=\lambda_\text{BB}\rho$.
This leaves us with three dimensionless essential parameters characterizing the
Bose polaron system \eqref{eq:H}: the mass ratio $\alpha=m/M$, the relative
interaction strength $\beta=\lambda_\text{IB}/\lambda_\text{BB}$
and a density parameter $\gamma=1/(\rho\xi)$.
$\lambda_\text{IB}$ is obtained from the scattering length $a_\text{IB}$
via $\lambda_\text{IB}=-\hbar/a_\text{IB} (1/M+1/m)$, and the
scattering length $a_\text{IB}$ is obtained from the parameters $U_0$
and $\sigma_U$ the Gaussian model interaction using the results
of Ref.~\onlinecite{jeszenszkiPRA18}.
We have confirmed the universality
of the interaction model, i.e.\ that our results
depend only on $\lambda_\text{IB}$ and not on the parameters $U_0$
and $\sigma_U$ if $\sigma_U$ is chosen very small. Too small values for $\sigma_U$ would
require a very fine discretization and correspondingly high numerical effort.
Therefore, we choose $\sigma_U=0.1$, where results differ only insignificantly from
the universal limit.

We compare results obtained with the inh-CP and the hom-CP equations to ensure numerical
consistency, and also to calculate the formation energy (called binding energy
in Ref.~\onlinecite{cucchiettiPRL06}) gained from self-localization if we do find self-localized polarons. But the main
goal of this work is to compare the inh-CP results and MF results,
i.e. results with and without including correlations,
to see whether self-localization still occurs
when impurity-boson correlations are included in the variational ansatz.
We note that both solving the hom-CP equation
and solving the MF equations is numerically straightforward
and fast since all quantities depend on a single coordinate, unlike
$f(x_0,x_1)$ in the inh-CP ansatz
\eqref{eq:Psi}.

\begin{figure}
\hspace*{-0.3cm}
\includegraphics[width=0.55\columnwidth]{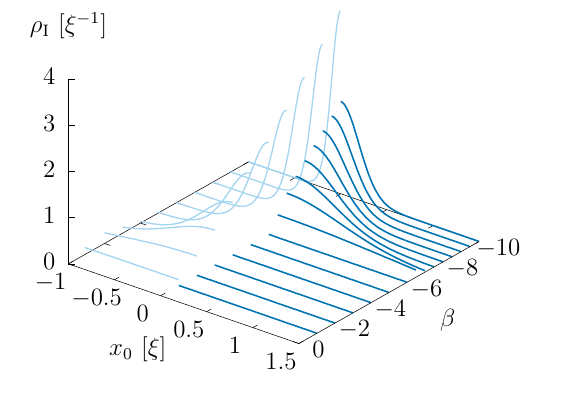}\hspace*{-0.4cm}
\includegraphics[width=0.55\columnwidth]{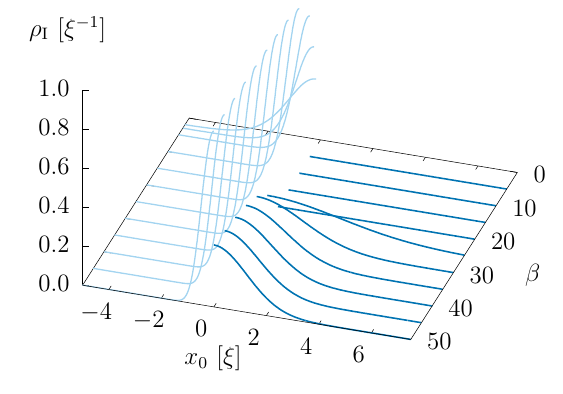}\\
\hspace*{-0.3cm}
\includegraphics[width=0.55\columnwidth]{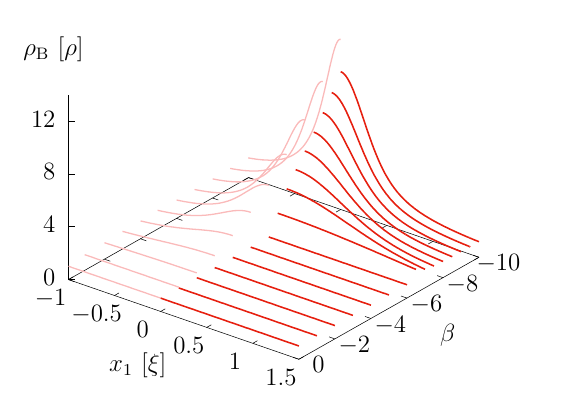}\hspace*{-0.4cm}
\includegraphics[width=0.55\columnwidth]{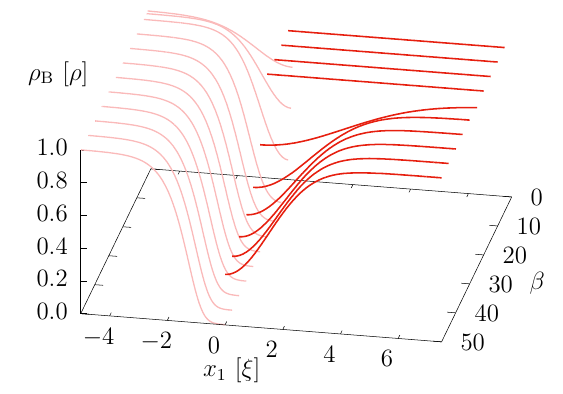}
\caption[]{\label{FIG:rhoI}
The impurity density $\rho_\text{I}(x_0)$ (top panels) and boson density $\rho_\text{B}(x_1)$ (bottom panels)
are shown as functions of $\beta$. The correlated polaron results are depicted on the positive side of
the coordinate axis $x_0$ or $x_1$, while the mean field results are depicted on the negative
side. The left and right panels show results for attractive and repulsive impurity-boson
interactions, respectively. A constant $\rho_\text{I}(x_0)$ and $\rho_\text{B}(x_1)$ means there
is no self-localization for the corresponding value of $\beta$. All results are for $\gamma=0.5$.
}
\end{figure}

In this work we restrict ourselves to equal impurity and boson mass, i.e.~$\alpha=1$.
The parameter $\gamma$ is related to the gas parameter, $\rho|a_\text{BB}|=2/\gamma^2$.
A small parameter $\gamma$ signifies weak boson-boson interactions (i.e.\ large $|a_\text{BB}|$)
and/or high density, while $\gamma\to\infty$ is the strongly correlated
Tonks-Girardeau limit~\cite{GIR60}. We study two cases, $\gamma=0.2$
and $\gamma=0.5$, which both correspond to a weakly interacting
Bose gas, where it may be justified to neglect boson-boson correlations
as done in the ansatz~\eqref{eq:Psi}. We vary the
relative impurity-boson interaction strength $\beta$ over a wide
range from strongly attractive to strongly repulsive.

\subsection{Density and localization length}

In Fig.~\ref{FIG:rhoI} we show the impurity density $\rho_\text{I}(x_0)$
(top panels) and the boson density $\rho_\text{B}(x_1)$ (bottom panels)
for attractive impurity-boson interactions, $-10\le\beta<0$, (left
panels) and repulsive interaction $0<\beta \le 50$ (right panels).
We show only half of the densities since they are assumed to be symmetric. The darker lines (positive coordinates) 
are the solutions of the inh-CP equations, while the lighter lines (negative coordinates)
are the solutions of the MF equations, calculated also in 
Ref.~\onlinecite{brudererEPL08}. All calculations in Fig.~\ref{FIG:rhoI} are done for $\gamma=0.5$.

The comparison in Fig.~\ref{FIG:rhoI} demonstrates that incorporating the impurity-boson correlations
strongly reduces the tendency towards self-localization. The MF approximation predicts
that the polaron self-localizes for {\it all} values of $\beta$, where
$\rho_\text{I}(x_0)$ and $\rho_\text{B}(x_1)$ becomes narrower for larger $|\beta|$.\cite{brudererEPL08}
Conversely, the ground state of the correlated polaron is qualitatively and quantitatively
quite different: for a wide $\beta$-range the polaron does not self-localize at all,
thus $\rho_\text{I}(x_0)$ and $\rho_\text{B}(x_1)$ are simply constant.
It may come as a surprise that especially for weak interactions the MF
approximation gives a wrong result regarding the question of self-localization,
which demonstrates that in one dimension correlations should never be neglected.
Only for sufficiently strong attraction or repulsion, the correlated polaron
self-localizes, but both $\rho_\text{I}(x_0)$ and $\rho_\text{B}(x_1)$ are significantly broader
than in the MF approximation.

\begin{figure}
\begin{center}
\includegraphics[width=1.03\columnwidth]{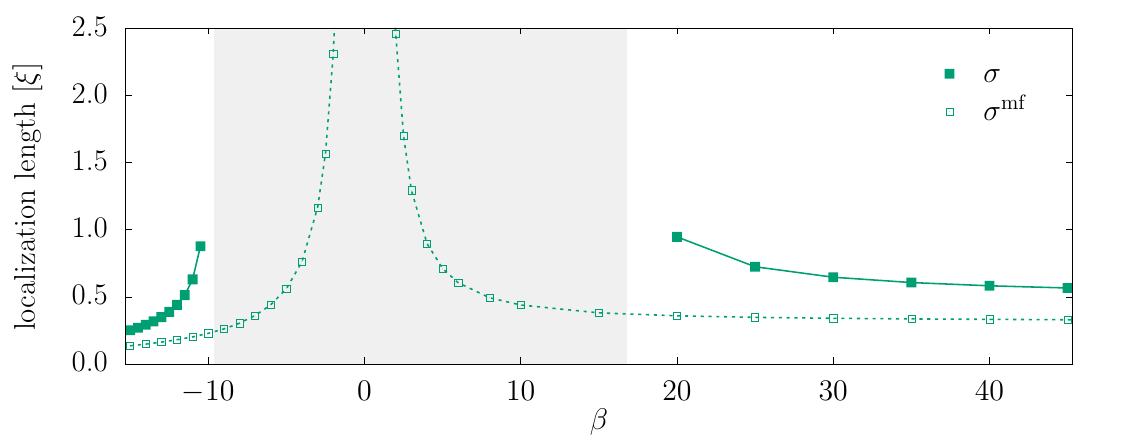}\\
\includegraphics[width=1.03\columnwidth]{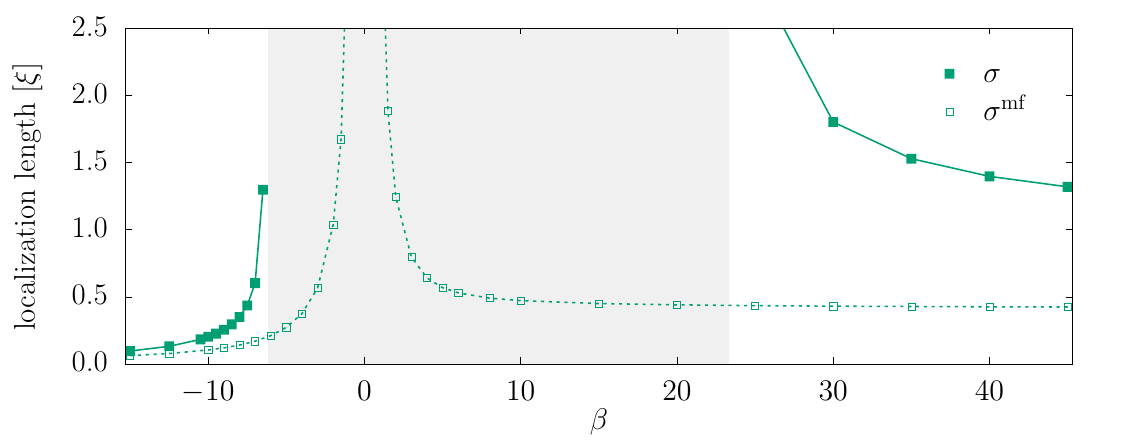}
\end{center}
\caption[]{\label{FIG:locallength}
The localization length $\sigma$ of a polaron is plotted as a function of $\beta$ for $\gamma=0.2$ (top panel) and 0.5 (bottom panel).
The filled and open symbols are the correlated and mean field results, respectively, the latter agreeing
with Ref.~\onlinecite{brudererEPL08}. The shaded area indicates the range of $\beta$ where no self-localization
occurs according to our correlated results.
}
\end{figure}

A localized polaron can be characterized by a localization length $\sigma$, e.g.\ by fitting a Gaussian
$\exp[-x_0^2/(2\sigma^2)]/(\sigma\sqrt{2\pi})$
to the impurity densities $\rho_\text{I}(x_0)$ shown in Fig.~\ref{FIG:rhoI}. $\sigma\to\infty$ means the polaron delocalizes.
In Fig.~\ref{FIG:locallength} we show the localization length $\sigma$ of the correlated polaron (filled squares)
and the corresponding $\sigma^\text{mf}$ of the MF polaron (open squares)
as functions of the relative interaction strength $\beta$ for $\gamma=0.2$ (top) and $\gamma=0.5$ (bottom).
Since in all our calculations, including the MF calculations,
we use a Gaussian interaction of finite width $\sigma_U=0.1$
instead of a contact potential, our results for
$\sigma^\text{mf}$ deviate slightly from Ref.~\onlinecite{brudererEPL08}, at most by 10\%.
Since the MF approximation predicts unconditional self-localization in 1D,
$\sigma^\text{mf}$ is finite for all $\beta\ne 0$. For the correlated polaron, we get a large
range of $\beta$ where the polaron is delocalized, indicated by the grey area. Therefore, $\sigma$ is not only significantly larger
than $\sigma^\text{mf}$, but it diverges at a critical attractive and repulsive relative interaction strength
$\beta_\text{cr,1}$ and $\beta_\text{cr,2}$, respectively, the value of which depends on $\gamma$.
Since a large $\sigma$ requires a large
computational domain, approaching the critical $\beta$ becomes numerically expensive and we estimate it by fitting to $a_1|\beta-\beta_\text{cr,1}|^{c_1}$ for the attractive side and
$a_2|1-\beta_\text{cr,1}/\beta|^{c_2}$ for the repulsive side (where $\sigma$ seems to saturate at a finite value for
large $\beta$). The estimates are tabulated in Tab.~\ref{table}.

\renewcommand{\arraystretch}{1.2}
\begin{table}
\begin{center}
\begin{tabular}{|c|cc||c|cc|}
\hline
$\gamma$ & $\beta_\text{cr,1}$ & $\beta_\text{cr,2}$ &
$\gamma_\text{P}$ & $\eta_\text{cr,2}$ & $\eta_\text{cr,2}$ \\
\hline
0.2 & -9.6 & 16.8 & 0.04 & -0.38 & 0.67 \\
0.5 & -6.2 & 23.3 & 0.25 & -1.55 & 5.82 \\
\hline
\end{tabular}
\end{center}
\caption[]{\label{table}
The critical relative interaction strengths $\beta_\text{cr,1/2}$ for self-localization, obtained from
solving the inh-CP equations. We also tabulate the results expressed
in alternative dimensionless units (see text) for better comparison with Ref.~\onlinecite{parisi_quantum_2017}.
}
\end{table}

The Bose polaron in one dimension was studied with diffusion Monte Carlo simulations
\cite{parisi_quantum_2017,grusdtNJP17}. The trial wave functions used in that work are translationally invariant, which may
mask a self-localization effect. Nonetheless, a relatively sharp increase of the polaron effective mass to a very large value was observed on both the attractive and repulsive side.
Parisi et al.~\cite{parisi_quantum_2017} considered equal masses for impurity and bosons, which allows
comparison with the present work. They
use the parameters $\gamma_\text{P}=\gamma^2$ and $\eta=\beta\gamma^2$ to characterize boson density/interactions and
impurity-boson interactions, respectively. For better comparison Tab.~\ref{table} provides the
critical interaction strength also in terms of $\gamma_\text{P}$ and $\eta$. The closest values of $\gamma_\text{P}$
compared to our values are $\gamma^\text{(MC)}_\text{P}=0.02$ and 0.2. Fig. 4 in Ref.~\onlinecite{parisi_quantum_2017} shows
that for $\gamma^\text{(MC)}_\text{P}=0.02$ the inverse effective mass essentially vanishes for $\eta\approx -1$ and for $\eta\approx 1$
for attractive and repulsive interactions, respectively; for $\gamma^\text{(MC)}_\text{P}=0.2$ the corresponding values
are $\eta\approx -2$ and $\eta\approx 10$, however the statistical fluctuations and the logarithmic scale makes it
hard to give precise numbers. Considering this uncertainty and our slightly different values for $\gamma_\text{P}$,
our prediction for the critical interaction strength for a self-localized polaron
ground state is consistent with that for an essentially infinite effective mass obtained with diffusion Monte Carlo.

\begin{figure}
\begin{center}
\includegraphics[height=0.47\columnwidth]{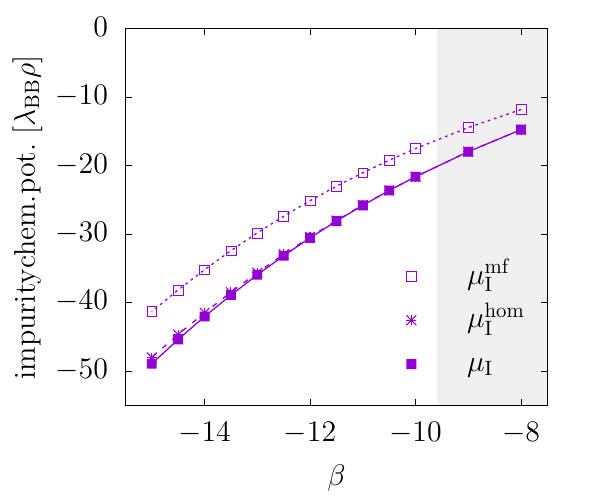}\hspace*{-0.03\columnwidth}
\includegraphics[height=0.47\columnwidth]{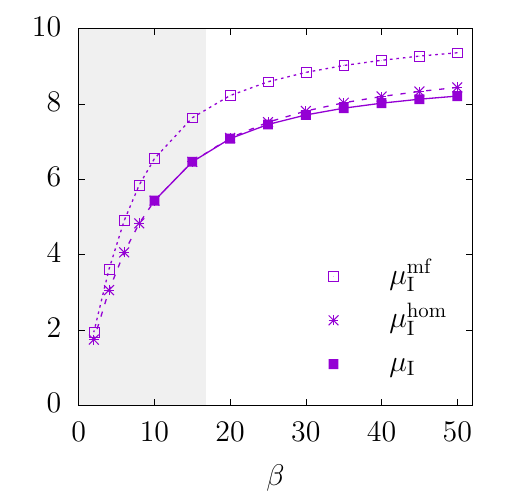}\\
\includegraphics[height=0.47\columnwidth]{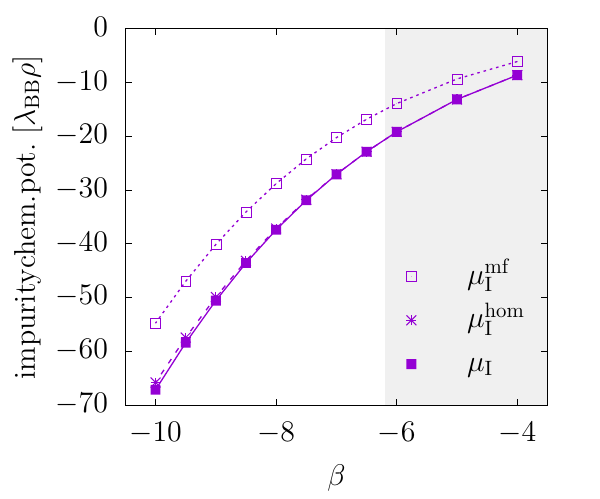}\hspace*{-0.03\columnwidth}
\includegraphics[height=0.47\columnwidth]{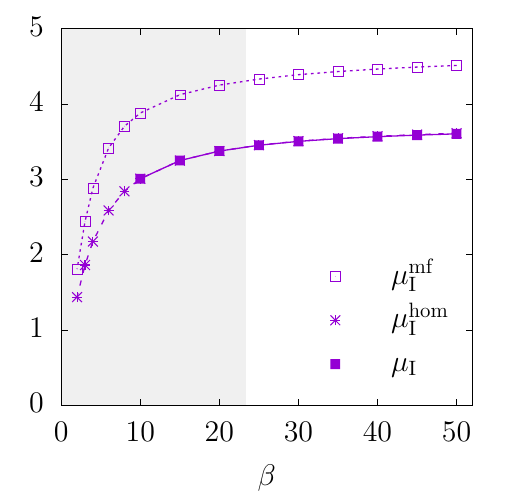}
\end{center}
\caption[]{\label{FIG:mu}
  The impurity chemical potential $\mu_\text{I}$ (filled squares) from the solution of the inhomogeneous correlated polaron
  equations is plotted as a function of $\beta$ for $\gamma=0.2$
  (top panels) and 0.5 (bottom panels), together with the mean field prediction $\mu_\text{I}^\text{mf}$ (open squares)
  and the homogeneous correlated polaron prediction $\mu_\text{I}^\text{hom}$ (stars). Left and right panels
  show attractive and repulsive impurity-boson interactions, respectively.
}
\end{figure}

\subsection{Chemical potential}

Solving the correlated polaron equations \eqref{Equ:var_eta_approx3} and \eqref{Equ:var_f_approx3} yields not only
$g(x_0)$ and $\tilde f(x_0,x_1)$ but also the impurity and boson chemical
potentials $\mu_\text{I}$ and $\mu_\text{B}$. For the latter we obtain the trivial result
$\mu_\text{B}/E_0=1$, i.e.\ the MF approximation of
the pure Bose gas, which is not altered by a single impurity in the thermodynamic limit.
Slight numerical deviations from unity provide a measure of finite size effects.

The impurity chemical potential $\mu_\text{I}$ provides nontrivial information.
According to the Ritz' variational principle, better variational wave functions yield lower energies,
closer to the exact ground state energy. This is also true for $\mu_\text{I}$, because it
is obtained by subtracting the constant $E_{0,N}$ from the ground state energy, see appendix~\ref{app1}.
Hence, the chemical potential of the correlated impurity must be lower than that of the MF impurity,
$\mu_\text{I}<\mu_\text{I}^\text{mf}$.  In Fig.~\ref{FIG:mu} we show $\mu_\text{I}$
and $\mu_\text{I}^\text{mf}$ as functions of $\beta$ for $\gamma=0.5$ (top panels)
and 0.2 (bottom panels). For all cases, $\mu_\text{I}^\text{mf}$ is higher than $\mu_\text{I}$, 
as it should be. Furthermore, we expect $\mu_\text{I}<0$ for $\beta<0$ and vice versa, which is
indeed the case for both $\mu_\text{I}$ and $\mu_\text{I}^\text{mf}$.
For attractive impurity-boson interactions, shown in the left panels,
$\mu_\text{I}$ shows no sign of saturating to a finite value when $\beta$ is decreased to stronger attraction,
in fact the slope steepens. For repulsive interactions (right panels), $\mu_\text{I}$ does saturate with
increasing $\beta$. This is consistent with the behavior of the localization
length shown in Fig.~\ref{FIG:locallength} for negative and positive $\beta$.

\begin{figure}
\begin{center}
\includegraphics[height=0.47\columnwidth]{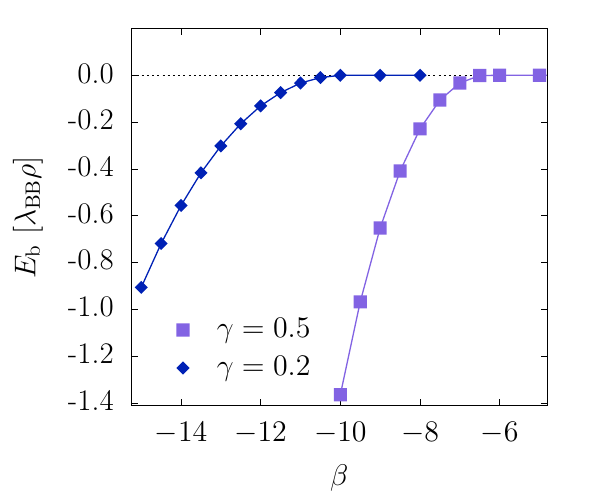}\hspace*{-0.03\columnwidth}
\includegraphics[height=0.47\columnwidth]{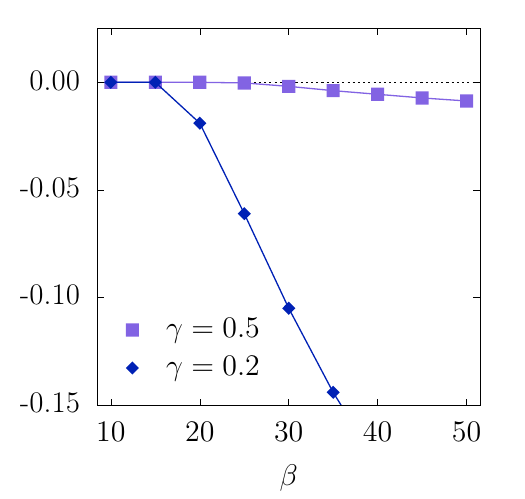}
\end{center}
\caption[]{\label{FIG:Eb}
  The formation energy $E_b=\mu_\text{I}-\mu_\text{I}^\text{hom}$ is plotted as a function of $\beta$, split into attractive
  and repulsive interaction (left and right panel). Self-localization happens only if $E_b<0$.
}
\end{figure}

The comparison between $\mu_\text{I}$ and $\mu_\text{I}^\text{mf}$ serves mainly as a check that we
did not converge to an unphysical local energy minimum. More interesting is the comparison of the chemical
potentials obtained from the inhomogeneous and the {\it homogeneous} polaron equations, $\mu_\text{I}$
and $\mu_\text{I}^\text{hom}$, respectively, because the difference is
the formation energy of self-localization, $E_b=\mu_\text{I}-\mu_\text{I}^\text{hom}$, i.e.\ the
energy gained by localization.
$\mu_\text{I}^\text{hom}$ is shown in Fig.~\ref{FIG:mu} together with $\mu_\text{I}$
and $\mu_\text{I}^\text{mf}$, but the difference between
$\mu_\text{I}$ and $\mu_\text{I}^\text{hom}$ is barely visible. In Fig.~\ref{FIG:Eb} we show the formation energy $E_b$,
which is about two orders of magnitude smaller than $\mu_\text{I}$, and its
determination without numerical bias is challenging. We note that the smallness of $E_b$
relative to $\mu_\text{I}$
would render its calculation by Monte Carlo simulation a formidable task.

If $\mu_\text{I}=\mu_\text{I}^\text{hom}$, thus $E_b=0$, no energy is gained from self-localization, which therefore does
not happen.  Indeed, in these cases the inh-CP solver converges to a constant
polaron density, $\rho_\text{I}=1/\Omega$, with the same correlation function $f(x_0,x_1)$
as that of the hom-CP solution, $f^\text{hom}(x_0-x_1)$.
If $\mu_\text{I}<\mu_\text{I}^\text{hom}$, thus $E_b<0$, self-localization lowers the
ground state with respect to a homogenous ground state. The critical
relative interaction strength $\beta_\text{cr,1}$ and $\beta_\text{cr,2}$
discussed above is just the point where $E_b$ becomes 0.

\begin{figure}
\begin{center}
\includegraphics[height=0.41\columnwidth]{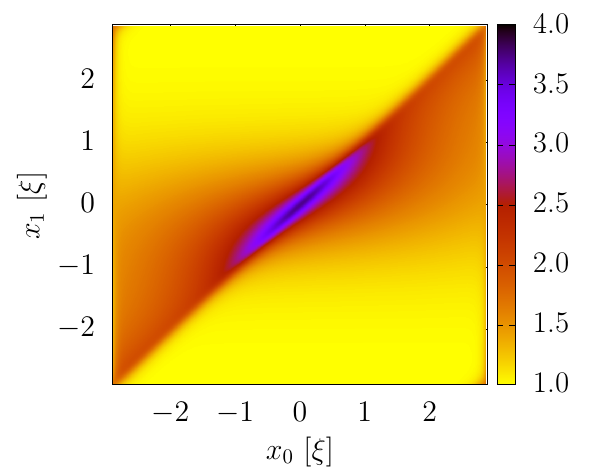}\hspace*{-0.02\columnwidth}
\includegraphics[height=0.41\columnwidth]{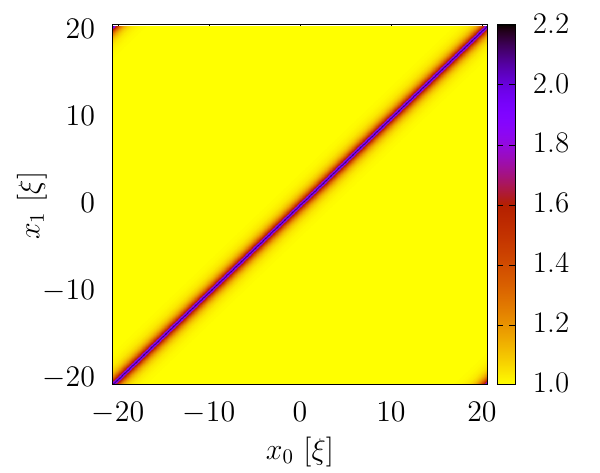}
\end{center}
\caption[]{\label{FIG:f}
Optimal pair correlation $f(x_0,x_1)$ obtained from solving the inhomogeneous correlated polaron
equations. For $\beta=-10$ (left panel) the polaron ground state is self-localized and
for $\beta=-5$ (right panel) the ground state is homogeneous. The values at the upper left and
lower right corners of the computational domain are a result of the periodic boundary conditions.
}
\end{figure}

We illustrate the difference between a homogeneous pair correlation
$f(x_0-x_1)$ of a delocalized ground state and the inhomogeneous pair
correlation $f(x_0,x_1)$ of a self-localized ground state
in Fig.~\ref{FIG:f} for $\gamma=0.5$. The left panel shows $f(x_0,x_1)$
for $\beta=-10$ (localized), which has only inversion symmetry.
The right panel shows $f(x_0,x_1)=f^\text{hom}(x_0-x_1)$
for $\beta=-5$ (homogeneous), which has translation symmetry with
respect to the center of mass $(x_0+x_1)/2$.

\section{Conclusions} 

We revisited the self-localization problem of an impurity in a Bose gas, where the mean field (MF)
approximation predicted self-localized polaron ground states in 3D,\cite{cucchiettiPRL06}, and later
in 2D and 1D \cite{brudererEPL08}; in particular in 1D, self-localization was predicted to happen for any strength of the impurity-boson interaction, quantified by the parameter $\beta$. Extending the
MF method using the Bogoliubov method
to account for quantum fluctuations has proven useful in many instances (dipolar interactions~\cite{limaPelsterPRA11},
self-bound Bose mixtures\cite{petrovPRL15}), but is still only a perturbative expansion.
In our work, we incorporate
optimized, inhomogeneous impurity-boson correlations in a non-perturbative way and derive
inhomogeneous correlated polaron (inh-CP) equations, which we solve
numerically for the 1D case. The results of this improved variational ansatz for the ground state wave function
shows that the MF approach is not sufficient to study polaron physics in 1D.
Impurity-boson correlations suppress the tendency towards
self-localization significantly, which happens only for strongly attractive or repulsive
impurity-boson interactions.  Despite being variational, our results are consistent with the sharp
increase of the effective mass of the polaron at a similar critical impurity-boson interaction strength
predicted by exact diffusion Monte Carlo simulations~\cite{parisi_quantum_2017}.

In case of the MF approximation, it is straightforward to see why it might predict a spurious self-localization
even for weak interactions:
without correlations, i.e. using a Hartree ansatz (\ref{eq:PsiMF}), a localized impurity density
and accordingly an inhomogeneous Bose density ``mimics'' the effect of a correlations as the most optimal
solution of the Ritz' variational problem.  For example, for repulsive interactions the Bose density
is suppressed around the localized impurity, lowering the total energy of a Hartree ansatz.
Instead, in a correlated many-body wave function like~\eqref{eq:Psi}, repulsion causes a correlation hole
in the pair distribution function, which does not require self-localization of the polaron.
Our method predicts self-localization only for strong impurity-boson interactions, but this is not
a rigorous proof that such a breaking of the translational invariance of the Hamiltonian~\eqref{eq:H}
is a real effect rather than a variational artifact.  Further refinements beyond the variational wave function~\eqref{eq:Psi},
such as boson-boson correlations or three body impurity-boson-boson correlations, may
push the transition to self-localization to even stronger interactions.
However, the above-mentioned consistency with exact Monte Carlo results lends credibility to the
correlated polaron ansatz \eqref{eq:Psi} in the regime of weak boson-boson interactions that we
studied in this work.

Experimental observation of a possibly self-localized polaron is challenging. The smallness of
the formation energy $E_b$ would require a low temperature, depending on the magnitude of $|\beta|$,
where strongly attractive interactions, $\beta<0$, are clearly favorable according to our results.
In higher dimensions, there is no evidence of a sharp increase of the effective mass of a polaron
three dimensions, according to quantum Monte Carlo simulations\cite{ardilaPRA15}, but
the MF approach~\cite{cucchiettiPRL06} does predict self-localization. Correlations tend to
be less important in higher dimensions, and the MF approach usually becomes a better approximation.
It will be interesting to see if there is a parameter regime where the correlated
polaron ansatz~\eqref{eq:Psi} is self-localized in more than one dimension. 
Furthermore, the inh-CP method can be generalized to time-dependent problems, similarly to
the time-dependent hypernetted-chain Euler-Lagrange method \cite{gartnerarxiv22}.
This allows to calculate the effective mass but also to study nonequilibrium dynamics of polarons after
a quench~\cite{grusdtPRA18}, such as an interaction quench of $\beta$.

Our results pertain only to neutral atomic impurities. For dipolar and especially ionic
impurities, which interact via long-ranged attractive
potentials with the surrounding Bose gas due to induced dipoles, the situation may be different. Ions
in BECs can dress themselves with a substantial cloud of bosons~\cite{astrakharchik_ionic_2021}, making
ionic polarons a more likely candidate for self-localization.

\begin{acknowledgments}
We thank Gregory Astrakharchik and David Miesbauer for fruitful discussions.
\end{acknowledgments}

\begin{appendix}

\section{Derivation of the inhomogeneous correlated polaron equations}\label{app1}

From the energy~\eqref{Equ:Efull2} and the resulting Lagrangian~\eqref{Equ:Lagrange_mu}
we derive the inh-CP equations~\eqref{Equ:var_eta_approx3} and \eqref{Equ:var_f_approx3}.
The first Euler-Lagrange equation \eqref{Equ:funcderiv_eta} becomes, after dividing by $2\bar{f}( x_0)^N$,
\begin{widetext}
\begin{align}
  \lambda \ \eta( x_0) =
&-\frac{\hbar^2}{2M} \ \Biggl\{ \frac{\partial^2 \eta( x_0)}{\partial  x_0^2}
+ 2 \ \frac{N}{\Omega} \ \frac{\partial \eta( x_0)}{\partial  x_0} \, \frac{1}{\bar{f}( x_0)} \int\! d x_1' \, f( x_0, x_1')\,
  \frac{\partial f( x_0, x_1')}{\partial  x_0} \nonumber\\
&\hspace{1.2cm}+ \frac{N}{\Omega} \, \frac{\eta( x_0)}{\bar{f}( x_0)} \int\! dx_1' \, f( x_0, x_1')\,
  \frac{\partial^2 f( x_0, x_1')}{\partial  x_0^2}
  + \frac{N(N-1)}{\Omega^2} \, \frac{\eta( x_0)}{\bar{f}( x_0)^{2}} \left[ \int\! d x_1' \, f( x_0, x_1')\,
  \frac{\partial f( x_0, x_1')}{\partial  x_0} \right]^2 \Biggr\} \nonumber\\
&+ \frac{\hbar^2}{2m} \frac{N}{\Omega} \, \frac{\eta( x_0)}{\bar{f}( x_0)} \int\! d x_1'
  \left( \frac{\partial f( x_0, x_1')}{\partial x_1'} \right)^2 
  + \frac{N}{\Omega} \, \frac{\eta( x_0)}{\bar{f}( x_0)} \int\! d x_1' \, f( x_0, x_1')^2 \ U( x_0, x_1') \nonumber\\
&+ \frac{\lambda_\text{BB}}{2} \ \frac{N(N-1)}{\Omega^2} \frac{\eta( x_0)}{\bar{f}( x_0)^{2}} \int\, d x_1' \, f( x_0, x_1')^4.
\label{Equ:var_eta_full}
\end{align}
Note that, when we multiply this equation by $\eta( x_0) \ \bar{f}( x_0)^{N}$
and integrate over $x_0$, we obtain $\lambda=E$, i.e.\ the Lagrange multiplier is indeed the energy.

Using \eqref{Equ:Efull2} and \eqref{Equ:Lagrange_mu}, the second Euler-Lagrange equation \eqref{Equ:funcderiv_eta} becomes,
after dividing by $\frac{2N}{\Omega}\bar{f}( x_0)^{N-1}$,
\begin{align}
& E \ \eta( x_0)^2 \ f( x_0, x_1) =
  -\frac{\hbar^2}{2M} \ \Biggl\{ \eta( x_0) \ \frac{\partial^2 \eta( x_0)}{\partial  x_0^2} \ f( x_0, x_1) 
  + \eta( x_0)^2 \ \frac{\partial^2 f( x_0, x_1)}{\partial  x_0^2} \nonumber\\
& \hspace{1.3cm} + 2 \ \frac{N-1}{\Omega} \, \frac{\eta( x_0)}{\bar{f}( x_0)^{-1}} \,  \frac{\partial \eta( x_0)}{\partial  x_0}\,
   f( x_0, x_1) \int\! d x_1' \, f( x_0, x_1') \, \frac{\partial f( x_0, x_1')}{\partial  x_0} \nonumber\\
& \hspace{1.3cm} + \frac{N-1}{\Omega} \, \frac{\eta( x_0)^2}{\bar{f}( x_0)} \, f( x_0, x_1) \int\! d x_1' \, f( x_0, x_1')\
  \frac{\partial^2 f( x_0, x_1')}{\partial x_0^2} \nonumber\\
& \hspace{1.3cm} + \frac{(N-1)(N-2)}{\Omega^2} \frac{\eta( x_0)^2}{\bar{f}( x_0)^{2}} \, f( x_0, x_1)
  \left[ \int\! d x_1' \, f( x_0, x_1') \, \frac{\partial f( x_0, x_1')}{\partial x_0} \right]^2
   + 2 \ \eta( x_0) \ \frac{\partial \eta( x_0)}{\partial  x_0} \ \frac{\partial f( x_0, x_1)}{\partial  x_0} \nonumber \\
& \hspace{1.3cm}+ 2 \ \frac{N-1}{\Omega} \frac{\eta( x_0)^2}{\bar{f}( x_0)} \, \frac{\partial f( x_0, x_1)}{\partial  x_0}
  \int\! d x_1' \, f( x_0, x_1') \, \frac{\partial f( x_0, x_1')}{\partial  x_0}  \Biggr\} \nonumber\\
& -\frac{\hbar^2}{2m} \, \Biggl\{ \eta( x_0)^2 \, \frac{\partial^2 f( x_0, x_1)}{\partial  x_1^2}
- \frac{N-1}{\Omega}\,
  \frac{\eta( x_0)^2}{\bar{f}( x_0)} \, f( x_0, x_1) \int\! d x_1' \, \left( \frac{\partial f( x_0, x_1')}{\partial x_1'} \right)^2 \Biggr\} \nonumber\\
&+ \eta( x_0)^2 \, f( x_0, x_1) \, U( x_0, x_1) 
+  \frac{\eta( x_0)^2}{\bar{f}( x_0)} \, f( x_0, x_1) \int\! d x_1' \, f( x_0, x_1')^2 \, U( x_0, x_1') \nonumber\\
&+ \lambda_\text{BB} \frac{N-1}{\Omega} \, \frac{eta( x_0)^2}{\bar{f}( x_0)} \, f( x_0, x_1)^3 
+ \frac{\lambda_\text{BB}}{2} \frac{(N-1)(N-2)}{\Omega^2} \, \frac{\eta( x_0)^2}{\bar{f}( x_0)^{2}} \, f( x_0, x_1) \int\! d x_1' \, f( x_0, x_1')^4.
\label{Equ:var_f_full}
\end{align}
We can simplify this lengthy equation by dividing by $\eta(x_0)$ and subtracting eq.~(\ref{Equ:var_eta_full}) multiplied by $f( x_0, x_1)$,
\begin{align}
  \Delta E\, \eta( x_0) \, f( x_0, x_1) =
& -\frac{\hbar^2}{2M} \, \Biggl\{ 2 \,\frac{\partial \eta( x_0)}{\partial  x_0} \, \frac{\partial f(x_0, x_1)}{\partial x_0}
+ 2 \frac{N-1}{\Omega} \, \frac{\eta( x_0)}{\bar{f}(x_0)} \, \frac{\partial f( x_0, x_1)}{\partial  x_0}
  \int\! d x_1' \, f( x_0, x_1') \, \frac{\partial f( x_0, x_1')}{\partial  x_0} \nonumber\\
& \hspace{1.3cm} + \eta( x_0) \ \frac{\partial^2 f( x_0, x_1)}{\partial  x_0^2} \Biggr\} \nonumber\\
& -\frac{\hbar^2}{2m} \, \eta( x_0) \, \frac{\partial^2 f(x_0, x_1)}{\partial x_1^2}
\ +\ \eta( x_0) \ U( x_0, x_1) \ f( x_0, x_1) \nonumber\\
  &+ \lambda_\text{BB} \frac{N-1}{\Omega} \, \frac{\eta( x_0)}{\bar{f}( x_0)} \, f( x_0, x_1)^2 \, f( x_0, x_1),
\label{Equ:var_f_full2}
\end{align}
where we abbreviate
\begin{align}
  \Delta E \equiv
& -\frac{\hbar^2}{2M} \, \Biggl\{ \frac{2}{\Omega} \, \frac{\partial \eta( x_0)}{\partial  x_0} \, 
  \frac{1}{\bar{f}( x_0)} \int\! d x_1' \, f( x_0, x_1') \, \frac{\partial f( x_0, x_1')}{\partial  x_0}
+ \frac{1}{\Omega} \, \frac{\eta( x_0)}{\bar{f}( x_0)} \int\! d x_1'\, f( x_0, x_1') \,
  \frac{\partial^2 f(x_0, x_1')}{\partial x_0^2} \nonumber\\
& \hspace{1.3cm} +2 \frac{N-1}{\Omega^2} \, \frac{\eta( x_0)}{\bar{f}( x_0)^{2}} \left[ \int\! d x_1'\, f( x_0, x_1') \,
  \frac{\partial f( x_0, x_1')}{\partial  x_0} \right]^2 \Biggr\} \nonumber\\
&+ \frac{\hbar^2}{2m} \frac{1}{\Omega} \, \frac{\eta( x_0)}{\bar{f}( x_0)} \int\! d x_1' \,
  \left( \frac{\partial f( x_0, x_1')}{\partial  x_1'} \right)^2 
  + \frac{1}{\Omega} \, \frac{\eta( x_0)}{\bar{f}( x_0)} \int\! d x_1' \, f( x_0, x_1')^2 \, U( x_0, x_1') \nonumber\\
&+ \lambda_\text{BB} \frac{N-1}{\Omega^2} \, \frac{\eta( x_0)}{\bar{f}( x_0)^{2}} \int\! d x_1' \, f( x_0, x_1')^4.
\end{align}
\end{widetext}
By comparison with eq.~\eqref{Equ:var_eta_full}, we see that $\Delta E$ is actually the difference between the energy $E=E_{1,N}$ of one impurity and $N$ bosons (see eq.~\eqref{Equ:Efull2}) and the energy $E_{1,N-1}$ of one impurity and $N-1$ bosons (see
eq.~\eqref{Equ:Efull2} with $N$ decremented by 1). Thus $\Delta E$ is the chemical potential
$\mu_\text{B}$ of the Bose gas,
\begin{align}
  \Delta E = E - E_{1,N-1} = \mu_\text{B}.
\end{align}

In the thermodynamic limit of an impurity in an infinitely large bath of bosons
we can simplify the equations \eqref{Equ:var_eta_full} and \eqref{Equ:var_f_full2} by letting
$N \rightarrow \infty$ and $\Omega \rightarrow \infty$,
with a constant boson density $\rho=\frac{N}{\Omega}$. This will provide a simple expression for $\bar{f}(x_0)$, eq.~\eqref{eq:barf}.
For large separation between the impurity and a boson, $\vert  x_0 -  x_1 \vert \rightarrow \infty$,
they are not correlated, $f( x_0,  x_1) \rightarrow 1$.  $h( x_0,  x_1) \equiv f( x_0,  x_1)^2 - 1$ provides
a measure for the correlations in the sense that $h\to 0$ means no correlations. We express $\bar{f}( x_0)$ in terms of $h$,
\begin{align*}
  \bar{f}( x_0)
  &= \frac{1}{\Omega} \int\! dx_1 \left[1+h(x_0,x_1)\right]\\
  &= 1 + \frac{1}{\Omega}\int\! dx_1 h(x_0,x_1)
  = 1 + \frac{\rho}{N} \int\! dx_1 h(x_0,x_1).
\end{align*}
Clearly, $\bar{f}( x_0) \rightarrow 1$ in the thermodynamic limit $N\to\infty$, but taken to the power of $N$, we obtain a
non-trivial function
\begin{align}
  \bar{f}( x_0)^N
  &= \left[ 1 + \frac{\rho}{N} \int\! dx_1 h(x_0,x_1) \right]^N\nonumber\\
  &\rightarrow\ \exp\left[ \rho \int\! dx_1 h(x_0,x_1) \right].
\label{Equ:barf_thermolim}
\end{align}
Most of the terms in eqns.~\eqref{Equ:var_eta_full} and \eqref{Equ:var_f_full2} are proportional to $\bar{f}( x_0)^{-1}$ or
$\bar{f}( x_0)^{-2}$, and one might be tempted to use $\bar{f}( x_0) \rightarrow 1$ in all of them.
However, the last term on the left side of eq.~\eqref{Equ:var_eta_full} requires closer attention.
With $N-1\approx N$ this term can be written as
\begin{equation}
  \frac{\lambda_\text{BB}}{2} \rho^2 \, \frac{\eta( x_0)}{\bar{f}( x_0)^{2}} \int\! d x_1' \, f(x_0,x_1')^4.
\label{Equ:termV}
\end{equation}
Because of $f(x_0,x_1)\to 1$ for $|x_0-x_1|\to\infty$, the integral scales with the volume $\Omega$, and we must
include corrections to $\bar{f}( x_0)^{-2}$ of order $1/\Omega$.
We expand $\bar{f}( x_0)=\Omega^{-1} \int d x_1' \ f( x_0, x_1')^2$ in powers of $\Omega^{-1}$ and obtain to first order
\begin{align}
	&\bar{f}( x_0)^{-2} \approx 1 - \frac{2}{\Omega} \int d x_1' \left( f( x_0, x_1')^2 -1 \right).
\end{align}
Thus, in the thermodynamic limit, the term~\eqref{Equ:termV} becomes
\begin{equation}
\frac{\lambda_\text{BB}}{2} \rho^2 \ \eta( x_0) \int d x_1' \ \Bigl( f( x_0, x_1')^4 - 2 \ f( x_0, x_1')^2 +2 \Bigr), \label{Equ:termV2}
\end{equation}
and Eq.~\eqref{Equ:var_eta_full} can be written
\begin{widetext}
\begin{align}
  E \, \eta( x_0) =
&-\frac{\hbar^2}{2M} \, \Biggl\{ \frac{\partial^2 \eta( x_0)}{\partial  x_0^2}
+ 2 \rho \, \frac{\partial \eta( x_0)}{\partial  x_0} \int\! d x_1' \, f( x_0, x_1') \, \frac{\partial f(x_0, x_1')}{\partial  x_0}
+ \rho \, \eta( x_0) \int\! d x_1' \, f( x_0, x_1') \, \frac{\partial^2 f( x_0, x_1')}{\partial  x_0^2} \nonumber\\
&\hspace{1.2cm}+ \rho^2 \, \eta( x_0) \left[ \int\! d x_1' \, f( x_0, x_1') \, \frac{\partial f( x_0, x_1')}{\partial  x_0} \right]^2 \Biggr\} \nonumber\\
&+ \frac{\hbar^2}{2m} \rho \ \eta( x_0) \int d x_1' \left( \frac{\partial f( x_0, x_1')}{\partial  x_1'} \right) ^2
+ \rho \, \eta( x_0) \int\! d x_1' \, f( x_0, x_1')^2 \, U( x_0, x_1') \nonumber\\
&+ \frac{\lambda_\text{BB}}{2} \rho^2 \, \eta( x_0) \int\! d x_1' \, \Bigl( f( x_0, x_1')^4 - 2 \, f( x_0, x_1')^2 +2 \Bigr).
\label{Equ:var_eta_approx}
\end{align}
\end{widetext}
Both sides of this equation scale linearly with $N$. Therefore, before taking the thermodynamic limit, we subtract
the MF energy of $N$ bosons without impurity $E_{0,N}=\frac{\rho^2}{2}\lambda_\text{BB}\Omega$ multiplied by $\eta( x_0)$.
With $E\equiv E_\text{1,N}$ we can then identify the impurity chemical potential $\mu_\text{I} = E_\text{1,N} - E_{0,N}$ on the
right-hand side of the resulting equation. Furthermore, we introduce the square root $g(x_0)=\sqrt{\rho_\text{I}(x_0)}$ of the impurity density
defined in eq.\eqref{eq:defrhoI}, which in the thermodynamic limit becomes, see
eq.\eqref{Equ:barf_thermolim},
\begin{align}
    \rho_\text{I}(x_0) &= \eta(x_0)^2 \bar{f}(x_0)^{N} \nonumber\\ &= \eta(x_0)^2\,\exp\left[ \rho\!\int\! d x_1' \ h( x_0,  x_1') \right].
    \label{eq:rhoIthermlimit}
\end{align}
This permits to write the one-body inh-CP equation in the final form
given in eq.~\eqref{Equ:var_eta_approx3}.

We use $N-1\approx N$ and $\bar{f}( x_0)\rightarrow1$ also in the two-body equation (\ref{Equ:var_f_full2}),
\begin{widetext}
\begin{align}
  \mu_\text{B} \ g( x_0) \ f( x_0, x_1) =
&-\frac{\hbar^2}{2M} \ \frac{1}{g( x_0)} \ \frac{\partial}{\partial  x_0} \ g( x_0)^2 \ \frac{\partial f( x_0, x_1)}{\partial  x_0}
-\frac{\hbar^2}{2m} \ g( x_0) \ \frac{\partial^2 f( x_0, x_1)}{\partial { x_1}^2} \nonumber\\
&+ g( x_0) \ U( x_0, x_1) \ f( x_0, x_1)
+ \lambda_\text{BB}\rho \ g( x_0) \ f( x_0, x_1)^2 \ f( x_0, x_1).
\label{Equ:var_f_approx2}
\end{align}
\end{widetext}
The final form \eqref{Equ:var_f_approx3} is obtained by defining $\tilde{f}( x_0, x_1) = g( x_0) f( x_0, x_1)$.

\section{Solving the correlated polaron equations}\label{app2}

The correlated polaron equations \eqref{Equ:var_eta_approx3} and \eqref{Equ:var_f_approx3}
are coupled nonlinear integro-differential equations for which we seek the
solution of lowest energy, according to the Ritz' variational principle. We need
a robust numerical scheme to obtain these solutions.

The one-body inh-CP equation \eqref{Equ:var_eta_approx3} has already the convenient
form of a nonlinear Schr\"odinger equations.
But the calculation of $\frac{1}{g(x_0)} \, \frac{\partial^2 g(x_0)}{\partial x_0^2}$ in
the effective potential~\eqref{Equ:Vf2} in the two-body inh-CP eq.\eqref{Equ:var_f_approx3}
can be numerically challenging: if $g(x_0)$ is self-localized, it decays
exponentially for large $x_0$. We therefore replace $\frac{1}{g(x_0)} \, \frac{\partial^2 g(x_0)}{\partial x_0^2}$
using the one-body equation~\eqref{Equ:var_eta_approx3} and obtain the alternative two-body equation
\begin{widetext}
\begin{align}
\left( \mu_\text{B} + \mu_\text{I} \right) \tilde{f}(x_0,x_1) =
&- \frac{\hbar^2}{2M} \ \frac{\partial^2 \tilde{f}( x_0, x_1)}{\partial { x_0}^2}
-\frac{\hbar^2}{2m} \ \frac{\partial^2 \tilde{f}( x_0, x_1)}{\partial { x_1}^2} 
+ \tilde V_f(x_0,x_1)\, \tilde{f}(x_0,x_1), \label{Equ:1D_f2_recombined}
\end{align}
with the effective two-body potential
\begin{align}
  \tilde V_f(x_0,x_1) &= V_g(x_0) +  U(x_0,x_1) + \lambda_\text{BB}\, \rho \, \frac{\tilde f(x_0,x_1)^2}{g(x_0)^2}.
  \label{Equ:Vf2}
\end{align}
\end{widetext}

Note that we still have to divide $\tilde f(x_0,x_1)$ by $g(x_0)$ for the calculation
of $V_g(x_0)$. This is the price for formulating the two-body equation as nonlinear
Schr\"odinger equation for $\tilde f(x_0,x_1)$. This
division by $g(x_0)$ can be problematic for localized solutions $g(x_0)$ if we choose the computation
domain too large.

Eqns.~\eqref{Equ:var_eta_approx3} and \eqref{Equ:1D_f2_recombined} are coupled
non-linear one- and two-body Schr\"odinger equations with
effective Hamiltonians $H_g=T_\text{I}+V_g$ and $H_f=T_\text{I}+T_\text{B}+\tilde{V}_f$, containing the
potentials \eqref{Equ:Vg} and \eqref{Equ:Vf2}, respectively.
We obtain the ground state by the imaginary time propagation.
We initialize $g$ and $f$ at
imaginary time $\tau=0$ with localized states, e.g.\ a MF solution, and then use small time steps $\Delta \tau$ together with
the Trotter approximation \cite{trotter59} to calculate an approximation of the
ground state by performing a large number $M$ of propagation steps until convergence
is reached:
\begin{align}
&g(M \Delta\tau) = \left( e^{-V_g/2 \, \Delta\tau} \ e^{-T_\text{I} \, \Delta\tau} \ e^{-V_g/2 \, \Delta\tau} \right)^M g(0)\\
&\tilde{f}(M \Delta\tau) = \left( e^{-\tilde{V}_f/2 \, \Delta\tau} \ e^{-(T_\text{I}+T_\text{B}) \, \Delta\tau} \ e^{-\tilde{V}_f/2 \, \Delta\tau} \right)^M \tilde{f}(0).
\label{Equ:SplitOperator}
\end{align}

In between time steps we have to normalize $g(x_0)$, which is the square root of the
impurity density,
\begin{equation}
\int dx_0 \ g(x_0)^2 = 1. \label{Equ:norm_g}
\end{equation}
Furthermore, in the thermodynamic limit the impurity and bosons should be uncorrelated for large separation, i.e.\ $f(x_0,x_1) \rightarrow 1$ for $\vert x_0-x_1\vert \rightarrow \infty$. In order to ensure this property, we specifically require
\begin{align}
&f(x_0=0,x_1\rightarrow\infty) = 1 \label{Equ:norm_f}.
\end{align}
In summary, we perform the following calculations
for each time step $\Delta\tau$ of the imaginary time propagation:
\begin{enumerate}
\item calculate $V_g$ \eqref{Equ:Vg} and $\tilde{V}_f$ \eqref{Equ:Vf2}
\item multiply $f$ by $g$ to get $\tilde{f}$
\item multiply $g$ by $\exp\bigl(-V_g/2 \ \Delta\tau \bigr)$
\item calculate the Fourier transform of $g$, multiply $g(k_0)$ by
 $\exp\bigl(-T_\text{I}(k_0) \Delta\tau \bigr)$ and transform back
\item multiply $g$ by $\exp\bigl(-V_g/2 \ \Delta\tau \bigr)$
\item normalize $g$ according to eq.~\eqref{Equ:norm_g}
\item multiply $\tilde{f}$ by $\exp\bigl(-\tilde{V}_f/2 \ \Delta\tau \bigr)$
\item calculate the Fourier transform of $\tilde{f}$, multiply $\tilde{f}(k_0,k_1)$
  by $\exp\bigl(- T_\text{I}(k_0)  \Delta\tau - T_\text{B}(k_1) \Delta\tau \bigr)$,
  and transform back
\item multiply $\tilde{f}$ by $\exp\bigl(-\tilde{V}_f/2 \, \Delta\tau \bigr)$
\item divide $\tilde{f}$ by $g$ to get $f$
\item normalize $f$ according to eq.~\eqref{Equ:norm_f}.
\end{enumerate}

In step 4 and 8,
$T_\text{I}(k_0)=\frac{\hbar^2}{2M} k_0^2$ and $T_\text{B}(k_1)=\frac{\hbar^2}{2m} k_1^2$ are the Fourier
transformed kinetic energies.

From the converged result, we calculate the impurity chemical potential $\mu_\text{I}$ using the change in
normalization by imaginary time propagation: we propagate one
time step without normalizing $g(x_0)$ and obtain $\mu_\text{I}$ from
\begin{equation}
\mu_\text{I} = - \frac{\ln \left( \int\! dx_0 \, g(x_0)^2 \right)}{2\Delta\tau}. \label{Equ:eigenenergy2}
\end{equation}


\end{appendix}

\bibliography{zotero,my,bec,ocshehy,papers}

\end{document}